\documentclass[10pt]{iopart}

\usepackage{graphicx}

\begin{document}

\title{Luminous infrared galaxies as possible sources of the UHE cosmic rays}
\author{A \'Smia\l{}kowski, M Giller, W Michalak}
\address{Division of Experimental Physics, University of Lodz, Pomorska
149/153, 90-236, Lodz, Poland}

\ead{asmial@kfd2.fic.uni.lodz.pl}
\submitto{\JPG}

\begin{abstract}
Ultra High Energy (UHE) particles coming from discrete extra\-galactic
sources are potential candidates for EAS events above a few tens of EeV.
In particular, galaxies with huge infrared luminosity triggered by collision and
merging processes are possible sites of UHECR acceleration. Here we check
whether this could be the case. Using the PSCz catalogue of IR galaxies we
calculate a large scale anisotropy of UHE protons originating in the population
of the luminous infrared galaxies (LIRGs). Small angle particle scattering in
weak irregular extragalactic magnetic fields as well as deflection by regular
Galactic field are taken into account. We give analytical formulae for
deflection angles with included energy losses on cosmic microwave background
(CMB). The hypotheses of the anisotropic and isotropic distributions of the
experimental data above 40 EeV from AGASA are check\-ed, using various
statistical tests. The tests applied for the large scale data distribution are
not conclusive in distinguishing between isotropy and our origin scenario for
the available small data sample. However, we show that on the basis of the small
scale clustering analysis there is a much better correlation of the UHECRs data
below GZK cut-off with the predictions of the LIRG origin than with those of
isotropy. We derive analytical formulae for a probability of a given
number of doublets, triplets and quadruplets for any density distribution of
independent events on the sky. The famous AGASA UHE triple event is found to be
very well correlated on the sky with the brightest extragalactic infrared source
within 70 Mpc - merger galaxies Arp 299 (NGC 3690 + IC 694). \end{abstract}
\section{Introduction}
The existence of the UHE cosmic rays, after their discovery almost half a
century ago, remains still puzzling. In advent of the new giant experiment named
in honour of Pierre Auger, there are many proposals to explain their origin.
UHE cosmic rays (UHECR) seem to be extragalactic because their arrival
directions are not correlated with the Galactic disk. Thus, sites of
their origin should be \emph{different} from normal galaxies (like our
Galaxy).
In this work we check the hypothesis that the powerful luminous infrared
galaxies (LIRGs) might be the UHECR sources.
Large fraction of bright IR galaxies are found to be interacting systems,
suggesting that collision and merging processes are mostly responsible in
triggering the huge IR light emission \cite{sa88}. Fraction of interacting
systems increases with IR luminosity and in the population of the most IR
luminous objects in the Universe almost all appear as gravitationally
interacting. Favourable environments for accelerating particles to UHE regime
via the first order Fermi process are provided by amplified magnetic fields on
the scale of tens kpc resulting from gravitational compression, as well as high
relative velocities of galaxies and/or superwinds from multiple supernovae
explosions \cite{c92,cp93}.
There have been several observational claims that colliding galaxies could
be possible sites of the UHECR origin. Al-Dargazelli \etal \cite{ad96} have
proposed that some clustering of UHECR shower directions above 10
EeV\footnote{$1EeV = 1\times 10^{18}$ eV} are associated with nearby colliding
galaxies. Uchihori \etal \cite{u00} have analysed combined world data from
Northern hemisphere experiments and concluded that two triplets and a doublet
lie in a vicinity of interacting systems. Takeda \etal \cite{t99} have noticed
that the interacting galaxy VV 141 at z=0.02 is a possible candidate for the
triplet of events above 50 EeV from the AGASA experiment. However, as we shall
show, there is another favourable candidate for the origin of this UHECR cluster
- Arp 299 (Mrk171, VV 118a/b), a member of LIRG class of extragalactic objects.
This system (RA$=171.4^{\circ}$, $\delta=58.8^{\circ}$), consisting of two
interacting starburst galaxies, is the closest extragalactic object (distance 42
Mpc for $H_{0}=75 km\,s^{-1}\,Mpc^{-1}$) with IR luminosity greater than
$5\times10^{11} L\odot$ ($\simeq 2\times10^{45}  ergs\,s^{-1}$) and it is the
brightest IR source within 70 Mpc. Such high IR luminosity is related to young
and violent star forming regions. There is also observational evidence of
superwind outflows, large scale strong radio emissions  and
the estimated supernova rate is about 0.6 per year \cite{ah00,he99,hy99,sl98}.
\\
\indent
The aim of the present paper is to check whether the LIRGs could be the
sites of origin of the UHECRs observed at the Earth.
Using the PSCz catalogue \cite{sa00} we construct the all-sky maps of UHE
proton intensities originating in LIRGs, taking into account effects of particle
propagation through the extragalactic medium and, as an
example, possible influence of the regular galactic magnetic field (GMF).
We check both hypotheses: origin in LIRGs and, on the other
hand, the isotropic distribution of the experimental data above 40 EeV from
AGASA \cite{h00}, using various statistical tests.
\section{Anisotropy calculations}
\subsection{The PSCz catalogue}
The PSCz catalogue consists of almost 15000 IR galaxies with known
redshifts, covering 84\% of the sky.
It should be noted that observational limitations cause that some of the
extragalactic objects may not have measured redshifts or may be
even unobserved in the dust obscured regions within the disk of the Galaxy
(the so called zone of avoidance).
Figure 1 (top) presents the distribution of all PSCz galaxies with known
redshifts. We can see patches on the sky regions excluded from the
catalogue. On inspection of the superimposed directions of the AGASA showers
above 40 EeV we can find only a few cases where experimental events lie in the
vicinity of excluded regions, which should not affect much our analysis. As the
potential UHECR sources we have selected objects with luminosities in the far
infrared (FIR) range exceeding LIRG limit $L_{FIR}=10^{11} L_{\odot}$, that is
over one order of magnitude greater than the estimated FIR luminosity of our
Galaxy \cite{cm89}, and with distances up to 1 Gpc, giving the total number
of 2811 sources (figure 1, bottom). A large fraction of such objects show to
be in an apparent stage of collision and merging.
\begin{figure}
 \includegraphics[width=12cm]{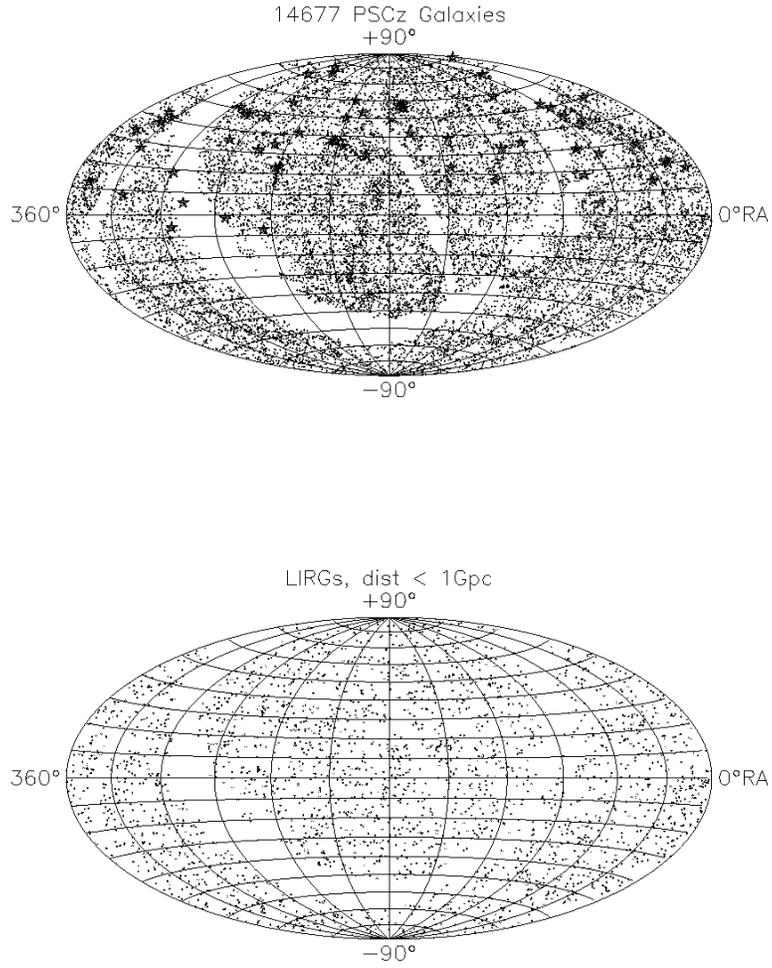}
 \caption{Top: Sky distribution of 14677 PSCz galaxies with known redshifts and
superimposed AGASA showers above 40 EeV (stars). Aitoff projection in
equatorial coordinates. Bottom: Selected 2811 objects with $L_{FIR}>10^{11}
L_{\odot}$ and with distances up to 1 Gpc}  \end{figure}
\subsection{CR propagation}
To predict the CR anisotropy expected in the model of LIRG origin we have
assumed the following:
\begin{itemize}
\item  Protons are injected at the sources with a spectrum $dN/dE\sim E^{-2}$
truncated at $10^{21}$ eV.
\item CR luminosity at the source is proportional to
its $L_{FIR}$ luminosity.
\item CR propagate through the intergalactic medium, where they are scattered
and suffer energy losses on the cosmic microwave background (CMB)\cite{bg88}.
\end{itemize}
To calculate CR arrival directions we have derived an analytical formula
for the distribution of the deflection angles, for the multiple small angle
scattering in weak irregular magnetic fields with continuous energy losses taken
into account (see Appendix A).
Since strength and structure of the extragalactic magnetic field are largely
unknown, we adopt here the upper limit for this field magnitude and coherence
scale: $B\,(l_{c})^{1/2}= 1\,nG\, (1Mpc)^{1/2}$, as measured by Faraday rotation
of radio signals from distant quasars \cite{k94}.
Because the analysis is strongly energy dependent, we consider
two energy regions 40 to 80 EeV and above 80 EeV, where this limit is just
about the Greisen-Zatsepin-Kuzmin (GZK) flux cut-off predicted from CR
interactions with the 2.7 K CMB radiation \cite{g66,zk66}.
\subsection{Maps of the expected anisotropy}
Equatorial maps of the expected intensity of the UHE protons originating in
LIRGs for two energy regions, 40-80 EeV and above 80 EeV, with sky coverage
and declination dependent exposure for the AGASA experiment \cite{t99}
(see also Appendix B), are presented in figures 2 (top) and 3 with superimposed
AGASA events.
\begin{figure}
 \includegraphics[width=12cm]{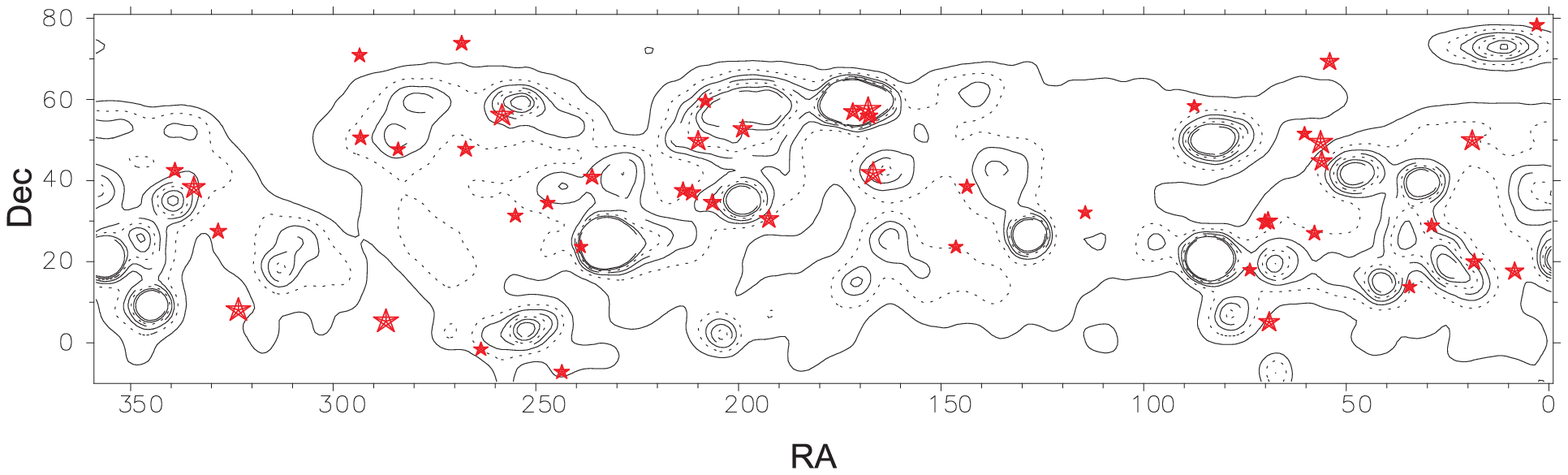}
 \includegraphics[width=12cm]{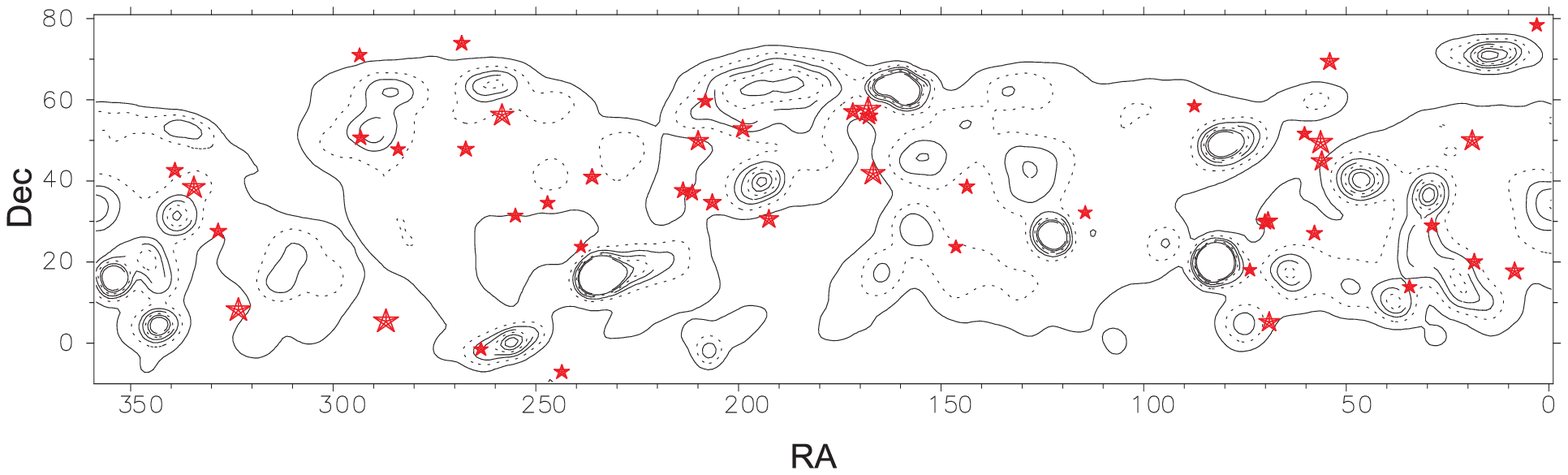}
 \caption{\label{1}Expected maps of proton intensities without
(top) and with
(bottom) influence of the regular galactic magnetic field (model
BSS-A) for protons with 40-80 EeV originating in LIRGs to be seen by
AGASA, with superimposed 47
AGASA shower directions in this energy range (stars scaled with
energy). Contours of constant flux per unit
solid angle are spaced linearly.}
\end{figure}
 \begin{figure}
 \includegraphics[width=12cm]{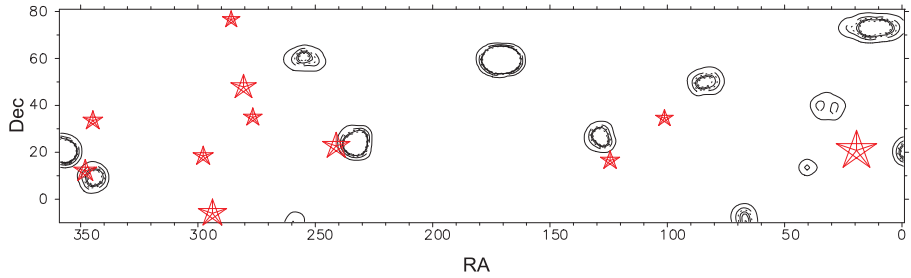}
 \caption{Expected map of proton intensities (energy above 80 EeV) from LIRGs
to be seen by AGASA, with superimposed 11 AGASA showers in this energy range
(no GMF).}
\end{figure}
In figure 2 (top) we can see that expected proton intensities for
40-80 EeV show good correlation with the distribution of the experimental
events from AGASA. Especially, clearly visible is the region of the sky where
high proton intensities from Arp 299 correlate with the AGASA triplet of events
(RA$\approx170^{\circ}$, $\delta\approx60^{\circ}$).
At higher energies, because of the rapid proton energy losses on CMB, only
sources located at distances not larger than 100 Mpc are visible on the map.
No correlation of the data events above 80 EeV with the expected CR
intensities can be seen (figure 3). There is an apparent group of UHE
showers in the region with RA$\approx300^{\circ}$, where no possible sources
exist. However some of the showers from this group lie close to the region
heavily obscured by the Galactic disk.
\subsection{Influence of Galactic magnetic field}
The global regular GMF structure is not well known. Analyses of
rotation measures of pulsars and extragalactic radio sources suggest that the
GMF has a bisymmetric spiral (BSS) form with field directions reversed from arm
to arm \cite{sf83,hq99}.
This is also supported by the observed value of the pitch angle of the
local field and number of field reversals within the Galactic disk
\cite{cc92,hq99,rl94}. To examine the influence of the regular
GMF on the extragalactic UHE fluxes we have chosen the bisymmetric spiral
model with field reversals and odd parity (BSS-A) adopted from
Stanev \cite{st97}.
Because the predictions are strongly model dependent, and the
procedure applied gives a rather simplified picture, the
analysis presented here is only an example of the possible extragalactic CR flux
distortion. It is presented for the lower energy range 40-80 EeV, where the
effect is stronger and easily visible.
\\
\indent
Simulations of a large number of monoenergetic
antiprotons ejected from the Earth and followed to the halo border through
GMF give flux modification factors depending on directions in the
'extragalactic' sky.
Then, for an assumed flux distribution of protons at the Galactic borders we
are able to calculate their fluxes at the Earth. Distortion effects i.e.
reduction or magnification and shifting of the particle fluxes are visible in
Figure 2 (bottom). The modification of the extragalactic UHECR intensities due
to the influence of the GMF BSS-A model seems to worsen the correlation with
the experimental data seen in the upper figure.
\section{Statistical tests and results}
\subsection{Smirnov-Cramer-von Mises test}
To check the hypotheses of isotropic and anisotropic distribution of the
experimental data above 40 EeV from AGASA, we have used Smirnov-Cramer-von
Mises (SCvM) free of binning test \cite{ea71}, modified for a 2-dimensional
distribution analysis. SCvM test is based on comparing the cumulative
distribution function $F(X)$ under hypothesis $H_{0}$ with the equivalent
function $S_{N}(X)$ of the data. The considered statistics $W^{2}$ is defined as
follows
\begin{equation}
W^{2}=\int_{-\infty}^{\infty}\Big(S_N(X)-F(X)\Big)^{2}f(X)dX
\end{equation}
where $f(X)$ is the probability density function corresponding to the
hypothesis $H_0$.  $S_N(X)$ is based on the experimental data (two
coordinates of the UHECR events on the equatorial map) and always
increases in steps of equal height, $N^{-1}$, where $N$ is the total number of
data. It is worth to note that this test is reliable even for small statistics
($N\geq3$).
Critical $NW^{2}$ values for the confidence level $\alpha=$0.1 and 0.05
are 0.347 and 0.461 respectively.
\begin{table}[ht]
\begin{indented}
\item[]\begin{tabular}{@{}lll}
\br
\textbf{E} &\textbf{Tested hypotheses} & \textbf{NW$^{2}$
values}
\\[0.5mm] \mr
40- &Isotropy & 0.063; 0.116; 0.202; 0.263 \\[1mm]
80 &LIRGs anisotropy & 0.069; 0.121; 0.106; 0.215 \\[1mm]
EeV &LIRGs anisotropy + GMF& 0.115; 0.195; 0.181; 0.247 \\[0.5mm] \mr
$>$80 &Isotropy & 0.042; 0.573; 0.658; 0.071  \\[1mm]
EeV &LIRGs anisotropy & 0.631; 0.351; 2.177; 0.338 \\[0.5mm] \br
\end{tabular}
\caption{ }
\end{indented}
\end{table}
\\
\indent
Each of the cumulative distribution functions, $F(X)$ and $S_{N}(X)$, is the
integral of the probability density function over the rectangle area defined by
the coordinate point and one of the corners of the map.
In this way, from the 2-dimensional maps we have constructed a 1-dimensional
$F(X)$ and $S_{N}(X)$, for four cases de\-pend\-ing on the chosen corner of the
map. Although the four obtained $NW^{2}$ values are not completely independent,
they provide a valuable insight into the analysis.
Results of tested hypotheses:
isotropy and LIRGs anisotropy (with and without GMF), energies
40-80 EeV and above 80 EeV are shown in table 1.
All the hypotheses in the energy range 40-80 EeV pass, assuming confidence
level $\alpha=$0.1 ($NW^{2}$=0.347). However, with the presence of the GMF
the agreement is worse. The hypotheses for energies above 80 EeV fail,
especially in LIRG anisotropy scenario.
\\
\indent
Irrespective of the above, table 1 shows that it is not easy to draw
conclusions from this test as the $NW^{2}$ values calculated for
different map corners scatter quite significantly.
Thus, it is advisable to apply another, hopefully more powerful,
statistical test
\subsection{Eigenvector test}
The idea is based on assigning unit directional vectors to the data points
on the celestial sphere.
By finding the normalized eigenvalues $\tau_{1}$, $\tau_{2}$, $\tau_{3}$ of the
orientation matrix $\mathbf{T}$ constructed for the N unit data vectors it is
possible to discriminate between the isotropic and some anisotropic
distributions \cite{fl93}.
Assuming $ 0\leq\tau_{1}\leq\tau_{2}\leq\tau_{3}\leq 1$
the empirical shape criterion $\gamma = $
$\left[log_{10}(\tau_{3}/\tau_{2})\right]/\left[log_{10}(\tau_{2}/
\tau_{1})\right]$,
and the strength parameter $\zeta=log_{10}(\tau_{3}/\tau_{1})$ are used
to discriminate girdle type from clustered distribution. Distributions of the
girdle and cluster type plot with $\gamma$ below and above unity, respectively.
Intermediate i.e. partly girdle, partly cluster distributions plot around the
line $\gamma=1$. Isotropic distributions plot with strength $\zeta$ near zero.
\\
\indent
 In figure 4 (left) there are shown distributions of 500 samples, consisting of
47 events each, simulated under isotropic (dots) and LIRGs anisotropic (crosses)
scenarios.
\begin{figure}
 \includegraphics[angle=270,width=12cm]{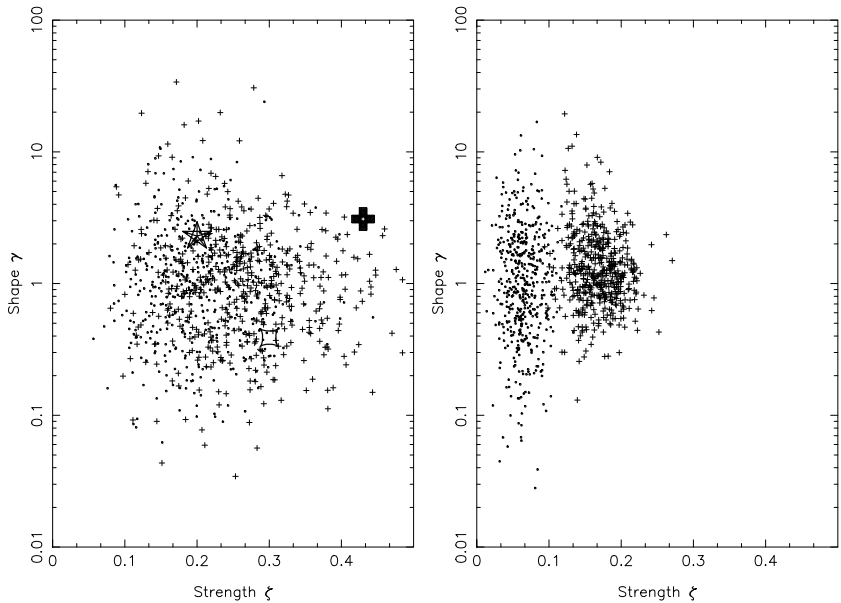}
 \caption{Left: Eigenvalue analysis distributions (each point
corresponds to 47 events) drawn from isotropy (dots)
and LIRGs anisotropy (crosses) scenarios. Points based on AGASA
data (star and thick cross represents respectively our and Medina Tanco
calculation from 'old' AGASA data - 47 events with energy above
40 EeV. The open square is for 'new' data - 47 events, 40-80 EeV); Right:
Simulations, as previously, but for large samples of 500 events each.}
\end{figure}
The open square denotes AGASA data, 47 events with energy 40-80 EeV.
The two hypotheses, LIRG anisotropy and isotropy, correspond to the two
regions in the $\zeta-\gamma$ plane which overlap significantly for the 47
event samples.
Mostly due to the small statistics considered, the eigenvector analysis is not
sensitive enough to distinguish between isotropic and LIRG anisotropic origin of
the experimental data.
Hopefully with a sample of more than 500 events, to be collected in a few months
of a full operational mode of the Auger observatory, the two distributions
should separate (figure 4, right).
\\
\indent
The eigenvector method has been already applied to the UHECR
anisotropy analysis by Medina Tanco \cite{mt01}.
The point corresponding to the 'old' AGASA data (47 events above 40 EeV),
calculated by this author, is shown on the $\zeta-\gamma$ plane (thick cross in
figure 4, left) lying well away from the simulated isotropic distribution.
However, the point calculated by us for the same AGASA data (star) has a lower
strength value and lies on the isotropy distribution.
Thus, contrary to the result of Medina Tanco, the AGASA data do not show such a
strong clustering (on the basis of the eigenvector analysis) to differ from the
isotropy distribution.
We have checked the correctness of our result by calculating the eigenvalues of
the orientation matrix also analytically (which is possible for a 3x3 matrix).
\\ \indent
We have not done the eigenvector analysis for the data above 80 EeV
because of the sample of 11 events was too small.
 \begin{figure}
 \includegraphics[angle=270,width=12cm]{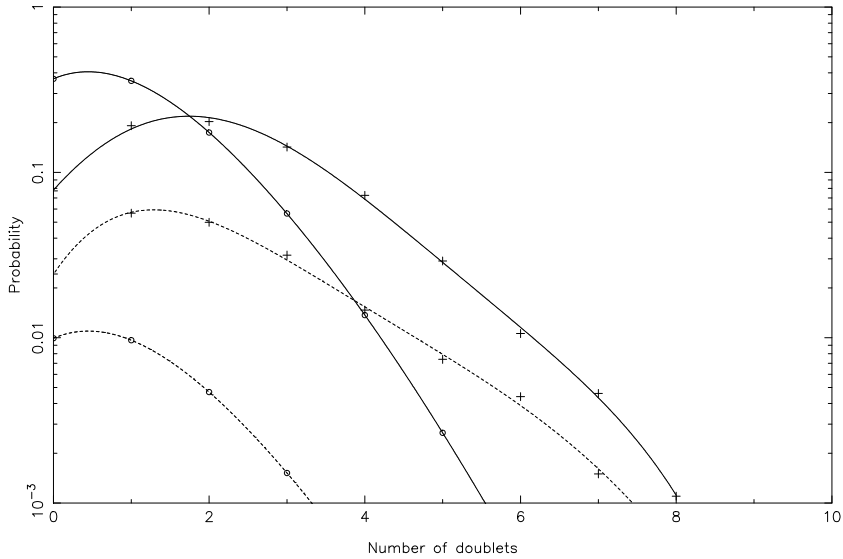}
 \caption{Probability distributions of doublets for isotropic
(circles) and
LIRGs anisotropic (crosses) scenarios for the cases of zero (solid
line) and
one triplet (dashed line).}
 \end{figure}
\subsection{Multiplet analysis}
Apart from analysing the large scale LIRG anisotropy (as in previous
chapters) we will check this hypothesis by analysing probabilities of
multiplets, i.e. groups of events with small angular separation.
\\
\indent
In the available\footnote{AGASA claims recently \cite{t01} three doublets in the
energy range (4-10)$\cdot 10^{19}$eV but it is not clear whether the energy of
one shower is below 8$\cdot10^{19}$ eV }
AGASA data from 40-80 EeV there are
two doublets (separation angles smaller than 2.5$^\circ$) and one triplet
(with criteria taken from the AGASA publication \cite{t99}, see also Appendix
B). Figure 5 shows probability distribution of the number of doublets based on
the $10^{4}$ samples of 47 events each, simulated from the LIRGs anisotropic map
(crosses) for the cases of zero and one triplet.
There are also similar probabilities for the isotropy scenario (circles), calculated
analytically (Appendix B).
From figure 5 we find that the probability of obtaining at least the observed
number of multiplets from LIRGs anisotropy scenario is $0.129$
($7.1\times10^{-2}$) for more than two (three) doublets, while from isotropy
(with the same assumed response in declination of the AGASA experiment) it is
only $0.67\times10^{-2} \, (0.2\times10^{-2}$). All AGASA events above 40 EeV
have one triplet and six doublets, a collection even less probable to be
obtained from isotropy.
\\
\indent
The above analysis may indicate only that there is some anisotropy
(which increases mainly the probability of triplet) but not necessarily that
correlated with LIRGs.
\begin{figure}  \includegraphics[angle=270,width=12cm]{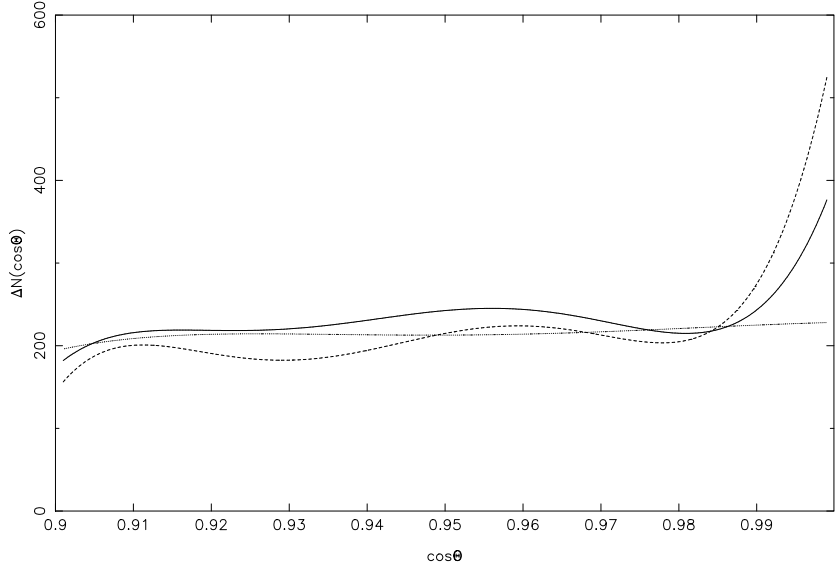}
\caption{Smoothed cos($\theta$) distributions for separation angles between
AGASA data (47 events 40-80 EeV) and 5000 events drawn from isotropic (dotted
line) and LIRG anisotropic (solid line) distributions. Dashed line
represents distribution of cosines within simulated LIRG events
(47$\times$5000) averaged over $10^{3}$ samplings. }
\end{figure}
A relationship between the set of the data and our hypothesis can be checked by
looking for small angle correlation between the simulated and the real events.
In figure 6 there are shown smoothed distributions of $cos\theta$ in the
range $0.9-1$, where $\theta$'s are separation angles between the AGASA events,
from one side (47 events, 40-80 EeV) and, from another side,
5000 events drawn from the isotropic (dotted line) or from the LIRG
(solid line) distributions (giving 47$\times$5000 cosine
values for each case). The obvious flat distribution of $cos\theta$ between the
AGASA data and isotropic events (dotted line) with a small rise from left to
right shows the effect of the AGASA declination efficiency.
The dashed line represents $cos\theta$ distribution for $\theta$ taken between
the events simulated from the LIRG map (47$\times$5000 events), averaged
over $10^{3}$ such runs.
Here, a significant rise in the range of a few degrees ($\theta<8$ deg,
$cos\theta>0.99$) indicates a clustering of the simulated LIRG events. A similar
rise is seen in the distribution of $cos\theta$ between AGASA
data and events simulated from LIRG (solid line). Taking into consideration the
sum of all cosines above 0.99, the resulting value lies within 1.67 $\sigma$
from the mean value obtained from the LIRG scenario, confirming quite a strong
correlation between the real and simulated LIRG events.
\\
\indent
In figure 7 there are presented distributions of 47 events from
AGASA and equivalent samples of events simulated from the map in
figure 2 (top). We have estimated that the probability of getting a doublet or a
triplet in the vicinity of Arp 299 is about 0.5. We can see that samples of the
47 events simulated from LIRGs anisotropy scenario, show good resemblance to the
experimental data distribution.
 \begin{figure}
 \includegraphics[angle=270,width=12cm]{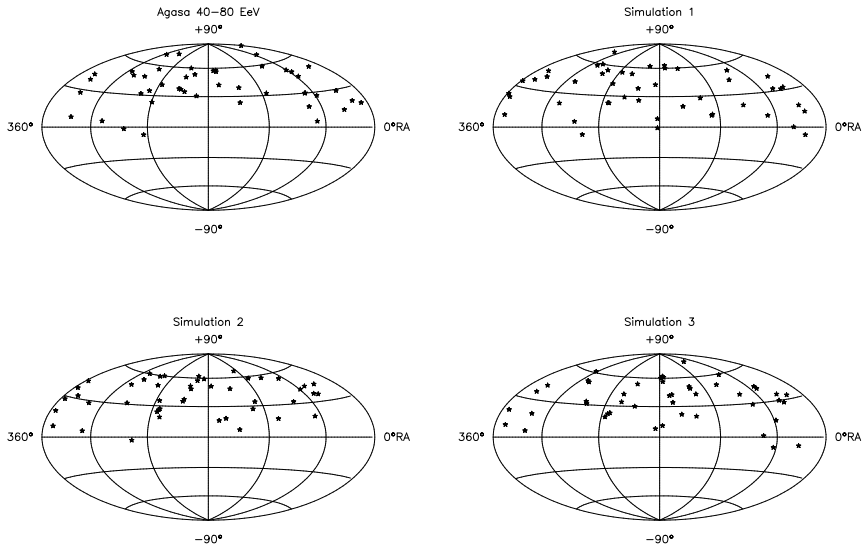}
 \caption{Sky distribution comparison of the 47 AGASA showers with energy
40-80 EeV (top left) and examples of three simulations from LIRG anisotropy
map. Equatorial coordinates.}
\end{figure}
\section{Astrophysical implications}
The strongest evidence for UHECRs origin may come from the correlation
between directions of the AGASA triplet of events with energies 53.5, 55.0 and
77.6 EeV and an energetic astrophysical source in the local
Universe. So far Arp 299 is the best candidate as a member of LIRGs, the
brightest infrared source within 70 Mpc and a system of colliding galaxies
showing intense, violent starburst activity at only 42 Mpc away. It should be
noted that Arp 299 appeared earlier as VV 118 in the list of candidates
for the AGASA triplet presented by Takeda \etal \cite{t99} but was not
recognized as a colliding, energetic system, and, as a result, was not given
enough attention. These authors point to another object VV 141 being a colliding
system at z=0.02. However, as Arp 299 is at a distance two times smaller and
fulfils the necessary criteria for CR acceleration, we think that it is this
object which could be the most probable CR source.
\\
\indent
Recent studies on the nature of LIRGs suggest that compact strong
radio emission may result from frequent multiple luminous radio
supernovae \cite{sl98}. These objects are a poorly known class of
supernovae with high nonthermal radio power indicating large kinetic energy
input to accelerate particles \cite{wb90}. In systems with intense starburst
activity in extremely dense molecular gas and strong magnetic field environment,
"nearly every supernova explosion results in a luminous radio SN with very high
radio power" \cite{sl98}.
It is also worth to note that there are some observations suggesting
a relationship  between gamma ray bursts (GRB) and supernovae explosions
\cite{l01}.
Very recently Weiler \etal \cite{wp01} have stated that GRB980425 and
radio loud supernova SN1998bw are possibly related. Thus, it is not
unreasonable to invoke here the intriguing hypothesis of the common origin of
the two most energetic, mysterious phenomena in the Universe, UHECRs and GRBs.
\section{Conclusions}
We have shown that the available data from AGASA in the energy range 40 to 80
EeV are not in contradiction with the expected anisotropy of CR produced in
LIRGs. After applying the GMF, the LIRG hypothesis passes as well, but an
apparent worse agreement with the data, suggests that the GMF
model used by us may not be appropriate. So, we may hope that, if point
sources existed they would be useful for determining the global structure
of the GMF. At energies above 80 EeV both the isotropic
and the LIRGs anisotropic distribution hypotheses are rejected. This
might be explained by an existence of a UHECR population of a different origin
above GZK cut-off.
\\
\indent
However, it seems that the SCvM test used here is not conclusive for
distinguishing between the isotropy and the LIRGs distributions.
Contrary to calculations done by other author, the eigenvector analysis has also
appeared unable to do this for the existing sample sizes.
Thus, we have considered a small scale clustering rather than a large
scale anisotropy.
Indeed, the analysis of multiplet probability gives good discrimination between
isotropic and anisotropic scenarios (figure 5).
The probability of  the occurring of two doublets and one triplet (three
doublets and one triplet) is 10 (20) times higher for the LIRG hypothesis than
that for isotropy . The analysis of the distributions of the
angular distances between the data and simulated LIRG events also
supports an existence of a correlation (figure 6).
The strongest argument for our hypothesis is the observation of the triplet from
the direction of Arp299, a system of colliding galaxies. We have obtained, via
simulations, a high probability of doublets and triplets from this source,
estimated to be about 0.5.
\\ \indent
Finally, with the prospect of new data to come from the Pierre Auger
Observatory in a few years, the arrival direction distribution of UHECR
should provide a much better clue to their origin.
\ack
The authors thank dr Jan Sroka for stimulating discussions.
This work was supported by: KBN (Polish State Committee for Scientific Research)
grant No 2PO3C 00618 and University of Lodz grant No 505/447.
\appendix
\section*{Appendix A. Small angle scattering}
\setcounter{section}{1}
Let us consider the statistical process of the small angle scattering of a
charged particle on $N$ randomly oriented magnetic cells. Here we derive
formulae for deviation angles and time delays on simple assumptions that
statistical variables $\delta\vec{n}_{k}$ (denoting change of unit vector
along direction of flight occurring in the $k$-th cell) are: 1) independent
and small, 2) cumulative change of direction is also small,
3) the number of cells $N$ is large.
 \begin{figure}
 \includegraphics[width=12cm,height=7.5cm]{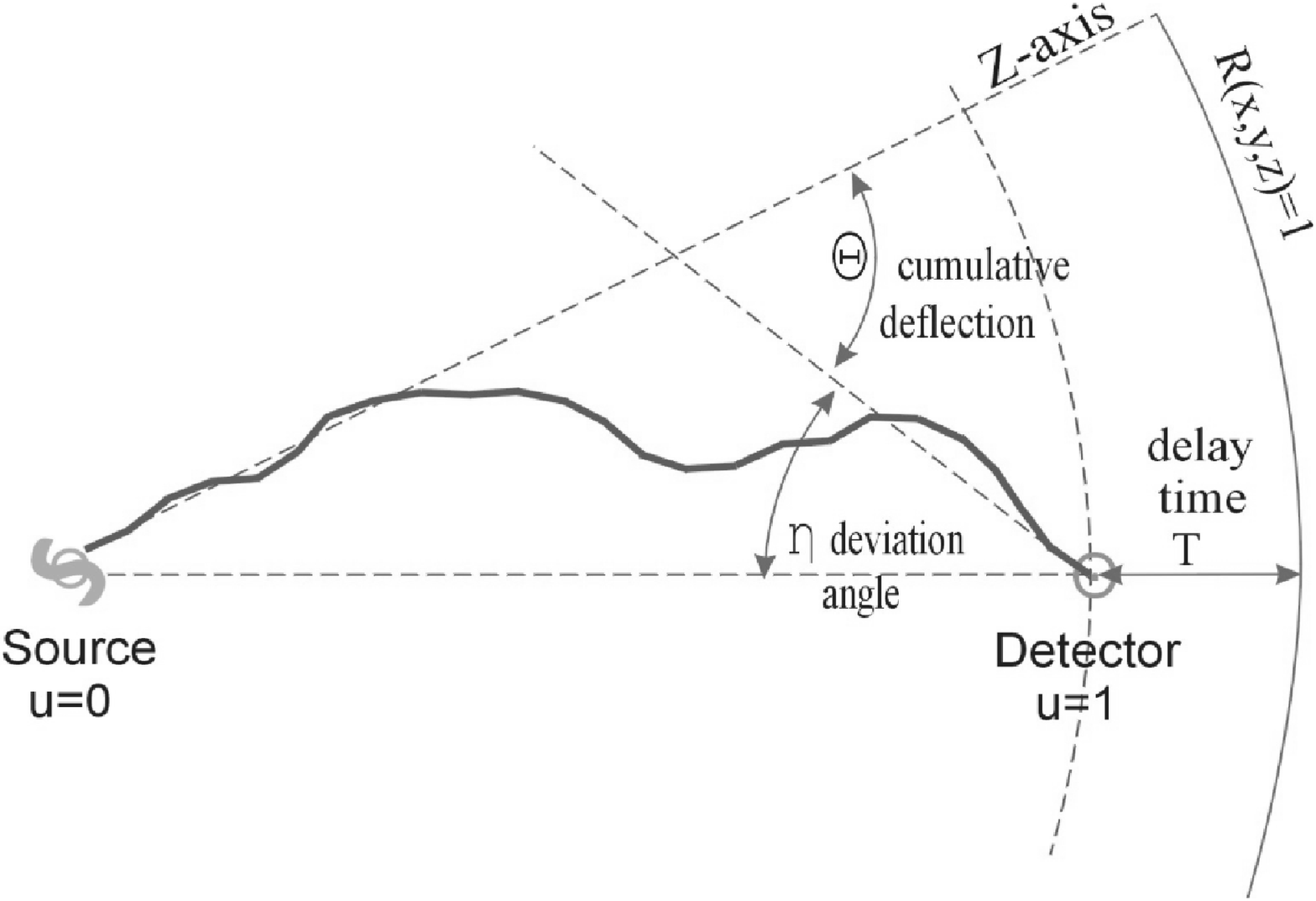}
 \caption{Schematic diagram of the particle trajectory through IGMF}
\end{figure}
As the particle energy changes along its trajectory, we have the relation
\begin{equation}
\langle\vec{\vartheta}_{k}^{2}\rangle=\langle\vec{\vartheta_{N}}^{2}
\rangle\cdot\Bigg(\frac{E_{N}}{{E}_{k}}\Bigg)^{2}\equiv
\langle\vec{\vartheta_{N}}^{2}
\rangle\cdot\Bigg(\frac{E_{u=1}}{E_{u}}\Bigg)^{2}
\end{equation}
Here $E_{N}$ and  $E_{k}$ denote particle energies in $N^{th}$ and
$k^{th}$ cells and $\langle\vec{\vartheta}_{N}^{2}\rangle$ is the mean square
scattering angle in the last $N_{th}$ cell, which is at the detector. We
also denote cells by a continuous variable $u=k/N$
instead of $k$ (i.e. $u=1$ indicates the detector). In this model the
particle's trajectory consists of $N$ segments of straight lines of equal length
$L=1/N$. If the source of particles is placed in the centre of a sphere of
radius $R=1$, then particle trajectories will end inside the sphere and the time
delay is just the additional time the particle needs to reach the sphere.
Looking back from the detector we miss the source by the deviation angle $\eta$
(figure A1). The formulae for the coordinates of the trajectory end: $x$,$y$,
$z$ ($z$-axis is the particle initial direction), time delay $T$ and angle
$\eta$ are shown below.
The cumulative change of direction is described by the angle $\Theta_{k}$ as:
$\vec{n_{k}}-\vec{n_{0}}=\sum_{i=1}^{k}\vec{\delta
n_{i}}=\sum_{i=1}^{k}
\vec{\vartheta_{i}}=\vec{\Theta_{k}}$.
\begin{eqnarray}
\lefteqn{x=\frac{1}{N}\cdot\sum_{k=1}^{N}\Theta_{xk};\quad
y=\frac{1}{N}\cdot\sum_{k=1}^{N}\Theta_{yk};\quad} \\
\nonumber
z=\frac{1}{N}\cdot\sum_{k=1}^{N}(1-
\frac{1}{2}\Theta^{2}_{xk}-\frac{1}{2}\Theta^{2}_{yk});\quad
\vec{\eta}=\sum_{k=1}{N}\frac{k}{N}\cdot\vec{\vartheta_{k}}
\end{eqnarray}

\begin{eqnarray}
\lefteqn{T\approx\Delta z-\frac{1}{2}\cdot(x^{2}+y^{2})\quad {} }
\nonumber\\
& & {}where \quad \Delta z=\frac{1}{N}\cdot
\sum_{k=1}^{N}(\frac{1}{2}\Theta^{2}_{xk}+\frac{1}{2}\Theta^{2}_{yk})
\end{eqnarray}
We can calculate statistical moments for the above random variables.
Here are some of them:
\begin{eqnarray}
\lefteqn{\langle\vec{\eta_{\, }}^{2}\rangle=N\cdot\langle\vec{\vartheta_{N}}^{2}
\rangle\cdot f_{12};\quad
\langle\vec{\eta}^{4}\rangle-\langle\vec{\eta}^{2}\rangle^{2}=\langle
\vec{\eta}^{2}\rangle^{2} {}}
\nonumber \\
& & {}\langle T\rangle=\frac{N}{2}\cdot\langle\vec{\vartheta_{N}}^{2}\rangle
\cdot(f_{2}-f_{11})
\end{eqnarray}
The constants $f_{11}$, $f_{12}$ and $f_{2}$ are defined by
the following integrals:
\begin{eqnarray}
\lefteqn{f_{11}=\int_{0}^{1}(1-u)^{2} A(u)\rmd u;\quad
f_{12}=\int_{0}^{1}u^{2} A(u)\rmd u; {}}
\nonumber \\
& & {}f_{2}=\int_{0}^{1}(1-u)\ A(u)\rmd u
\end{eqnarray}
where
\begin{equation}
A(u)=\Bigg(\frac{E_{u=1}}{E_{u}}
\Bigg)^{2}\nonumber
\end{equation}is the particle energy relation from source to detector
assuming continuous energy losses. It should be noted that allowing for
fluctuations in the energy losses would give larger deflection angles.
The values of the derived moments have been positively checked in simulations with
various functions $A(u)$.
\appendix
\section*{Appendix B. Probability of multiplets.}
\setcounter{section}{2}
Here we derive analytical formulae for average numbers of doublet, triplet and
quadruplet events on the assumption that the expected flux is known for any
direction and single showers occur independently of each other.
The last assumption means that a particular number of multiplets of a given kind (e.g. that
of doublets) undergoes Poisson distribution with the expected value equal to the average of
the expected (local) values over the entire map.
Let us denote by  $\lambda(\delta,\alpha)$  ($\delta$ - declination, $\alpha$ - RA) the
expected angular density of showers, i.e. the expected number of showers per unit solid
angle within the measurement time ($\lambda=const$ for isotropy). The actual number of
registered showers depends, of course, on the efficiency $\eta(\delta,\alpha)$ of the particular
air shower array.
Assuming that only showers with zenith angles $\Theta$ smaller than some $\Theta_{z}$ are
considered, we define here that $\eta(\delta,\alpha)=\langle f(t)cos(\Theta(t))\rangle$ where
$f(t)=1$ while $\Theta(\delta,\alpha,t)<\Theta_{z}$ and $f(t)=0$ otherwise. The brackets mean
time average. It can be derived that:
\begin{equation}
\eta(\delta,\alpha)=\bigl[\sin(\varphi)\sin(\delta)\alpha_{z}+
\cos(\varphi)\cos(\delta)\sin(\alpha_{z})\bigr]/\pi
\end{equation}
where the angle $\alpha_{z}$ is  defined by:
\begin{equation}
\cos(\alpha_{z})=\frac{\cos(\Theta_{z})-\sin(\varphi)\sin(\delta)}
{\cos(\varphi)\cos(\delta)}
\end{equation}
This formula agrees well with the experimental efficiency of the AGASA experiment \cite{t99}.
The expected angular density $\rho(\delta,\alpha)$ of registered showers equals
$\eta(\delta,\alpha)\cdot\lambda(\delta,\alpha)$.
If the mean number of showers detected is N, then we have the normalizing
relation $N=\int\rho(\delta,\alpha)\rmd\Omega$ and the average flux $\lambda$ is
determined:
\begin{equation}
\lambda=\frac{N}{\int\eta(\delta,\alpha)\rmd\Omega}
\end{equation}
 \begin{figure}
 \includegraphics[width=12cm]{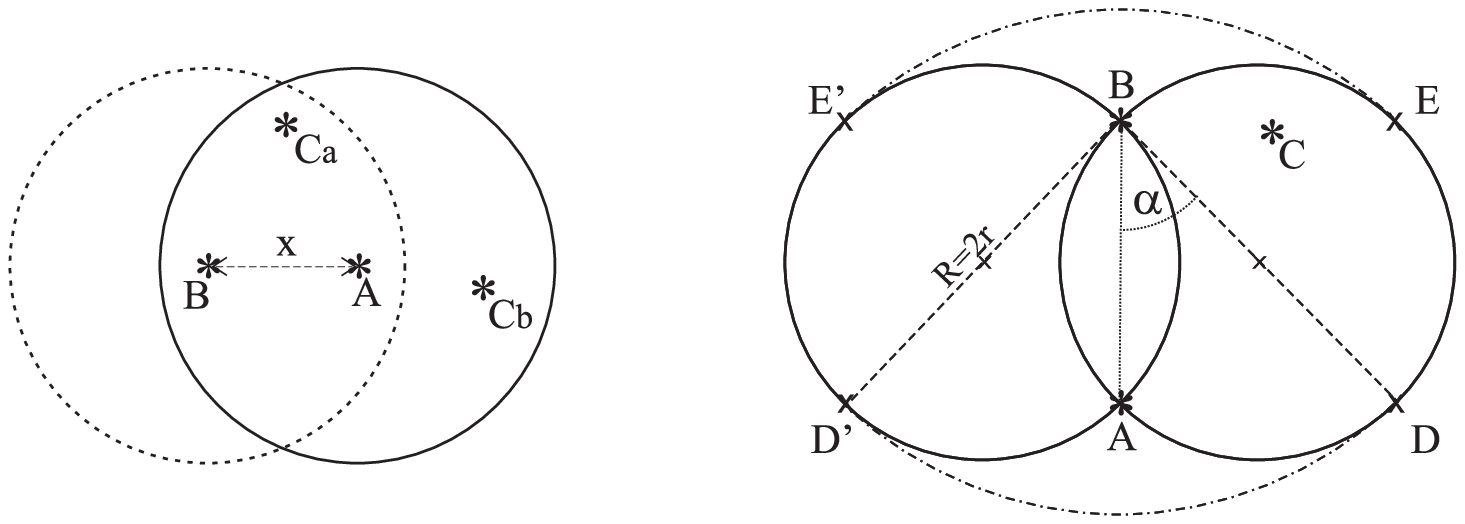}
 \caption{Diagrams for triplet description under 1st and 2nd criterion}
\end{figure}
We assume that the number of registered showers in an experiment is large enough
to adopt it as the average number of showers which would have been registered by
the detector in many such runs.
The criterion for classifying two showers as a doublet is that the
angle between them should be smaller than some given small value $R$.
As the expected number of showers within the small angle $R$ around a given
direction is $\mu=\rho(\pi R^{2})$ (assuming constant $\rho$ within
the circle), the local detection density of doublets can be written as
$\rho_{2}=\frac{1}{2}\rho\mu \rme^{-\mu}$ . The factor $\frac{1}{2}$ stands for
the fact that both members  of a doublet count as one doublet.
Of course, the local detection density of singles, where there is no
companion within the angle $R$ from the shower considered is
$\rho_{1}=\rho\rme^{-\mu}$.
\\
\indent
The situation is more complex for the case of a triplet. Two criteria may be
distinguished:
\begin{description}
\item[1st] - another two showers should be closer than $R$ to the shower
considered (in the centre) - criterion applied by AGASA.
\item[2nd] - three showers should be within a circle with diameter $R$.
\end{description}
The 1st criterion is illustrated in the figure B1 (left). In the centre of the
right circle there is the
shower denoted as A and within the radius R there are another two showers,
out of which the shower denoted as B is in a distance $x$ from the
first one, in the centre of  the left circle. The third shower may be either on
the area where the two circles overlap (Ca) or outside it (Cb). In the first
case around all three showers a triplet will be found (the same). In the second
case the triplet is found only around shower A.
\\
Let S(x) denote the ratio of overlapping area on the figure to the area of the
circle. We have the following relations:
\begin{equation}
S(x)=\frac{R^{2}\bigl[2\alpha-\sin(2\alpha)\bigr]}{\pi R^{2}}
\end{equation}
where $\cos(\alpha)=x/(2R)$. The average value of the ratio $S(x)$ equals:
\begin{equation}
\langle
S(x)\rangle=2\int_{0}^{R}S(x)\frac{x}{R^{2}}\rmd x=1-\frac{3\sqrt{3}}{4\pi}
\end{equation}
Thus, the local detection density of triplets under 1st criterion can be written
as :
\begin{eqnarray}
\lefteqn{ \rho_{3}=\rho\Bigl[\frac{1}{3}\langle S(x)\rangle
+\Big(1-\langle S(x)\rangle \Big)\Bigr]
\frac{\mu^{2}}{2!}\rme^{-\mu} \quad {} }
\nonumber \\
& & {}
\quad =\frac{1}{3}\rho\Bigg(1+\frac{3\sqrt{3}}{2\pi}\Bigg)
\frac{\mu^{2}}{2!}\rme^{-\mu}
\end{eqnarray}
The mean numbers of doublets $N_{2}$ and triplets $N_{3}$ detected from all directions
are found by integrating their detection densities over the whole solid angle.
\\
We get:
\begin{eqnarray}
\lefteqn{
N_{2}=\int\rho_{2}(\delta,\alpha)\rmd\Omega=\frac{1}{2}(\pi
R^{2})\int(\lambda\eta)^{2}
\rme^{-\mu}\rmd\Omega\approx\frac{1}{2}(\pi R^{2})\lambda^{2}\int\eta^{2}\rmd\Omega
}
\nonumber \\
& & {}
N_{3}=\int\rho_{3}(\delta,\alpha)\rmd\Omega=\frac{1}{6}(\pi R^{2})^{2}
\Bigg(1+\frac{3\sqrt{3}}{2\pi}\Bigg)\int(\lambda\eta)^{3}\rme^{-\mu}\rmd\Omega
\nonumber \\
& & {}
\quad \approx
\frac{1}{6}(\pi R^{2})^{2}\lambda^{3}\Bigg(1+\frac{3\sqrt{3}}{2\pi}\Bigg)
\int\eta^{3}\rmd\Omega
\end{eqnarray}
where we have assumed that $\lambda=const$.
Similar consideration as those for detection of triplets under the 1st criterion
lead us to the corresponding formulae for quadruplets under the same criterion:
\begin{eqnarray}
\lefteqn{ \rho_{4}=\rho\Bigl[\frac{1}{4} S_{34}
+\frac{1}{2}S_{14}+\big(1- S_{34}-S_{14}\big)\Bigr]
\frac{\mu^{3}}{3!}\rme^{-\mu} \quad {} } \nonumber \\
& & {}
N_{4}=\int\rho_{4}(\delta,\alpha)\rmd\Omega=\frac{1}{6}(\pi
R^{2})^{3}\lambda^{4}\cdot 0.66 \cdot
\int\eta^{4}\rme^{-\mu}\rmd\Omega
\end{eqnarray}
where $S_{34}$ is the probability that three showers randomly distributed within
a circle with a forth shower in the centre will constitute the same quadruplet
if \emph{any} shower out of the three is chosen as the central one.
The probability that only \emph{one} shower out of the three, chosen as the
central one, makes the same quadruplet possible is denoted $S_{14}$. It has
been found by integration that $S_{34}=0.274$ and $S_{14}=0.270$.
\\ \indent
The formulae for detection densities $\rho_{2},\rho_{3},\rho_{4}$ derived above
allow for such shower configuration, where the shower B in the figure B1 (left)
being a member of a doublet or a triplet is at the same time a member of a
triplet or a quadruplet, respectively. After excluding such situations we get
for doublets the following formula:
\begin{eqnarray}
\lefteqn{ \rho^{'}_{2}=\rho_{2}-\rho\cdot 3\Big(1-\langle S(x)\rangle\Big)
\frac{\mu^{2}}{2!}\rme^{-\mu} \quad {} }
\end{eqnarray}
Figure B1 (right) illustrates the situation for classifying
three showers as a triplet for the 2nd criterion, i.e. three showers have to be
within a circle with diameter $R$. The two circles there have diameter $2r=R$.
Two showers from directions A and B are at the angular distance AB $< R$. To
find the area where the third shower should be we draw the circle of radius
$R/2$ through the point A and rotate it around point A until point B is inside
the circle, so that points E goes to point E'. Similarly, we rotate the circle
around point B until point A is inside the circle (point D goes to D'). To form
a triplet the third shower must be located within area E'EDD'.
\\ \indent
Let $S_{2}(x)$ be the ratio of the framed area to $\pi R^{2} $ for a given
value AB$=x$. It follows that:
\begin{equation}
S_{2}(x)=\big[6\alpha+\pi-\sin(2\alpha)\big]\frac{1}{4\pi} \quad where \quad
\cos(\alpha)=\frac{x}{R}
\end{equation}
Taking into account the distribution of distances AB we can calculate the
average value of the $S_{2}(x)$:
\begin{equation} \langle
S_{2}(x)\rangle=2\int_{0}^{R}S_{2}(x)\frac{x}{R^{2}}\rmd x =\frac{9}{16}
\end{equation}
Further considerations for the local detection density of
triplets under the 2nd criterion are only for the case $\mu\ll 1$.
So, in a similar way, we get that:
\begin{eqnarray}
\lefteqn {
\rho_{3}=\rho\frac{1}{3}\langle
S_{2}(x)\rangle\frac{\mu^{2}}{2!}=\frac{3}{32}\rho\mu^{2}
{} }
\\
& & {}
N_{3}=\int\rho_{3}(\delta,\alpha)\rmd\Omega=\frac{1}{6}(\pi
R^{2})^{2}\frac{9}{16}\int(\lambda\eta)^{3}\rmd\Omega=\frac{3}{32}(\pi
R^{2})^{2}\lambda^{3}\int\eta^{3}\rmd\Omega
\nonumber
\end{eqnarray}
Numbers of observed doublets, triplets, quadruplets etc undergo Poisson
distributions with the corresponding mean (expected) values $ N_{2},
N_{3},\,  (N^{'}_{2}, N^{'}_{3})$, etc as calculated above.
The formulae for the mean numbers of the multiplets have been positively checked
in simulations. For the AGASA experiment the following values apply: \\
$\Theta_{z} =45^{\circ}; \varphi=35^{\circ}47'; R =2.5^{\circ}$ \\
We find that :\\
$\int\eta\rmd\Omega=\pi(1-\cos^{2}\Theta_{z})=1.57 \,sr;\,
\int\eta^{2}\rmd\Omega=0.38 \,sr; \,\int\eta^{3}\rmd\Omega=0.097 \,sr;
\,\int\eta^{4}\rmd\Omega=0.0255 \,sr$ \\ For the expected value of showers N we
use the actual number (registered in the energy range 40 to 80 EeV) - 47 showers
and obtained that: \\ $N_{2}=1.018;\,
N_{3}=0.0283;\,N^{'}_{2}=0.954;\,N^{'}_{3}=0.0268$ under 1st criterion;
$N_{3}=0.0087$ under 2nd criterion; $N_{4}=4.57\times10^{-4}$ for quadruplets
under 1st criterion.
\\ \indent
Thus the probability of obtaining two doublets or
more and one triplet or more, for the isotropic sky without point sources equals
$\bigl[ 1-exp(-N^{'}_{2})-N^{'}_{2}\cdot exp(-N^{'}_{2})\bigr]\cdot
\bigl[1-exp(-N_{3})\bigr]=0.247\cdot 0.0279=6.9\times 10^{-3}$. We see that it
is mainly the triplet event that makes the isotropy hypothesis very unlikely.

\end{document}